\documentclass[twocolumn]{aastex631}
\usepackage{graphicx}
\usepackage{natbib}
\usepackage{latexsym}
\usepackage{amssymb}
\usepackage{comment}
\usepackage{amsmath}
\usepackage{url}

\def\msun{\ifmmode {\rm\,M_\odot}\else ${\rm\,M_\odot}$\fi}
\def\Msun{\ifmmode {\rm\,\it{M_\odot}}\else ${\rm\,M_\odot}$\fi}
\def\lsun{\ifmmode {\rm\,L_\odot}\else ${\rm\,L_\odot}$\fi}
\def\Lsun{\ifmmode {\rm\,\it{L_\odot}}\else ${\rm\,L_\odot}$\fi}
\def\rsun{\ifmmode {\rm\,R_\odot}\else ${\rm\,R_\odot}$\fi}
\def\Rsun{\ifmmode {\rm\,\it{R_\odot}}\else ${\rm\,R_\odot}$\fi}
\def\Tsun{\ifmmode {\rm\,T_\odot}\else ${\rm\,T_\odot}$\fi}
\def\arcsec{\ifmmode {^{\prime\prime}}\else $^{\prime\prime}$\fi}
\def\asec{\ifmmode {^{\prime\prime}}\else $^{\prime\prime}$\fi}
\def\arcmin{\ifmmode {^{\prime}}\else $^{\prime}$\fi}
\def\amin{\ifmmode {^{\prime}}\else $^{\prime}$\fi}

\begin{document}

\title{The Colorado Ultraviolet Transit Experiment ($CUTE$) Mission Overview}

\author[0000-0002-1002-3674]{Kevin France}
\affiliation{Laboratory for Atmospheric and Space Physics, University of Colorado Boulder, Boulder, CO 80303}

\author[0000-0002-2129-0292]{Brian Fleming}
\affiliation{Laboratory for Atmospheric and Space Physics, University of Colorado Boulder, Boulder, CO 80303}

\author[0000-0002-4701-8916]{Arika Egan}
\affiliation{Laboratory for Atmospheric and Space Physics, University of Colorado Boulder, Boulder, CO 80303}

\author[0000-0002-0875-8401]{Jean-Michel Desert}
\affiliation{Anton Pannekoek Institute for Astronomy, University of Amsterdam, P.O. Box 94249, Noord Holland, NL-1090GE Amsterdam, the Netherlands}

\author[0000-0003-4426-9530]{Luca Fossati}
\affiliation{Space Research Institute, Austrian Academy of Sciences, Schmiedlstrasse 6, 8042 Graz, Austria}

\author[0000-0003-3071-8358]{Tommi T. Koskinen}
\affiliation{Lunar and Planetary Laboratory, University of Arizona, Tucson, AZ 85721--0092}

\author[0000-0001-7131-7978]{Nicholas Nell}
\affiliation{Laboratory for Atmospheric and Space Physics, University of Colorado Boulder, Boulder, CO 80303}

\author[0000-0001-7624-9222]{Pascal Petit}
\affiliation{Institut de Recherche en Astrophysique et Plan\'etologie, Universit\'e de Toulouse, CNRS, CNES, 14 avenue Edouard Belin, 31400 Toulouse, France}

\author[0000-0001-5371-2675]{Aline A. Vidotto}
\affiliation{Leiden Observatory, Leiden University, PO Box 9513, 2300 RA, Leiden, the Netherlands}

\author{Matthew Beasley}
\affiliation{Southwest Research Institute, Boulder, CO 80302}
\author{Nicholas DeCicco}
\affiliation{Laboratory for Atmospheric and Space Physics, University of Colorado Boulder, Boulder, CO 80303}

\author[0000-0002-4166-4263]{Aickara Gopinathan Sreejith}
\affiliation{Laboratory for Atmospheric and Space Physics, University of Colorado Boulder, Boulder, CO 80303}
\affiliation{Space Research Institute, Austrian Academy of Sciences, Schmiedlstrasse 6, 8042 Graz, Austria}

\author{Ambily Suresh}
\affiliation{Laboratory for Atmospheric and Space Physics, University of Colorado Boulder, Boulder, CO 80303}

\author{Jared Baumert}
\affiliation{Laboratory for Atmospheric and Space Physics, University of Colorado Boulder, Boulder, CO 80303}

\author[0000-0001-9207-0564]{P. Wilson Cauley}
\affiliation{Laboratory for Atmospheric and Space Physics, University of Colorado Boulder, Boulder, CO 80303}

\author[0000-0003-1701-7143]{Carolina Villarreal D'Angelo}
\affiliation{Instituto de Astronom\'ia Te\'orica y Experimental, CONICET-UNC. Laprida 854, C\`ordoba, Argentina }

\author[0000-0002-8636-3309]{Keri Hoadley}
\affiliation{The University of Iowa, Dept. of Physics \& Astronomy, Van Allen Hall, Iowa City, IA 52242}

\author{Robert Kane}
\affiliation{Laboratory for Atmospheric and Space Physics, University of Colorado Boulder, Boulder, CO 80303}
\affiliation{Blue Canyon Technologies 2550 Crescent Dr, Lafayette, CO 80026}

\author{Richard Kohnert}
\affiliation{Laboratory for Atmospheric and Space Physics, University of Colorado Boulder, Boulder, CO 80303}

\author{Julian Lambert}
\affiliation{Laboratory for Atmospheric and Space Physics, University of Colorado Boulder, Boulder, CO 80303}

\author{Stefan Ulrich}
\affiliation{Laboratory for Atmospheric and Space Physics, University of Colorado Boulder, Boulder, CO 80303}

\correspondingauthor{Kevin France}
\email{kevin.france@colorado.edu}

\begin{abstract} 
Atmospheric escape is a fundamental process that affects the structure, composition, and evolution of many planets.  The signatures of escape are detectable on close-in, gaseous exoplanets orbiting bright stars, owing to the high levels of extreme-ultraviolet irradiation from their parent stars. The {\it Colorado Ultraviolet Transit Experiment} ($CUTE$) is a CubeSat mission designed to take advantage of the near-ultraviolet stellar brightness distribution to conduct a survey of the extended atmospheres of nearby close-in planets. The $CUTE$ payload is a magnifying NUV (2479~--~3306~\AA) spectrograph fed by a rectangular Cassegrain telescope (206mm $\times$ 84mm); the spectrogram is recorded on a back-illuminated, UV-enhanced CCD.  The science payload is integrated into a 6U Blue Canyon Technology XB1 bus. $CUTE$ was launched into a polar, low-Earth orbit on 27 September 2021 and has been conducting this transit spectroscopy survey following an on-orbit commissioning period.  This paper presents the mission motivation, development path, and demonstrates the potential for small satellites to conduct this type of science by presenting initial on-orbit science observations. The primary science mission is being conducted in 2022~--~2023, with a publicly available data archive coming on line in 2023.
\end{abstract}


\section{INTRODUCTION}
The history of observational astronomy has been marked by the push to ever larger and more capable telescopes and instruments.  The 2010s witnessed the development of a new generation of large astronomical observatories.  Both on the ground and in space, facilities 2~--~3 times the primary mirror diameter of the previous state-of-the-art were brought closer to fruition for implementation in the 2020s, 2030s, and 2040s, including the {\it James Webb Space Telescope}~\citep{gardner06,rigby22}, thirty-meter class ground-based telescopes (e.g., \citealt{30m_ovr}), and advanced ultraviolet/optical (UV/O) facilities such as the Large Ultraviolet/ Optical/ Infrared Surveyor (LUVOIR; LUVOIR Final Report 2019).   The large mission studies conducted ahead of the 2020 Decadal Survey on Astronomy and Astrophysics drove the recommendation for NASA’s suite of Future Great Observatories, a series of probe- and flagship-class missions offering many order-of-magnitude gains in the scientific grasp across numerous areas of astrophysics.  In parallel with this large observatory development, numerous small telescope arrays have come on-line or have been expanded, and NASA’s science divisions made significant new investments in small satellites covering a range of scientific topics.   

Small telescopes at ground-based sites have excelled at detecting and characterizing new objects in the time-variable sky, including supernovae eruptions~\citep{dong16} and tidal disruption events~\citep{hammerstein22} from the Zwicky Transient Facility~\citep{ztf_ovr} and All Sky Automated Survey for SuperNovae (ASAS-SN)~\citep{asassn_ovr}.  The impact of small telescopes has also been powerful for the detection of extrasolar planets, including many Jovian-sized planets from Wide-Angle Search for Planets (WASP)~\citep{wasp_ovr} and the Kilodegree Extremely Little Telescope (KELT)~\citep{kelt_ovr}, and some of the most promising rocky planets for study with $JWST$ from the MEarth~\citep{charbonneau09}, Transiting Planets and Planetesimals Small Telescope (TRAPPIST)~\citep{gillon11}, and Search for habitable Planets EClipsing ULtra-cOOl Stars (SPECULOOS)~\citep{burdanov18} facilities.   

The recent decadal survey  highlighted the power of small space-based telescopes, astronomical CubeSats and smallsats, for ``monitoring of sources for weeks or months at time, and at wavelengths not accessible from the ground'', complementing the {\it Hubble Space Telescope}’s surveys in areas of transmission spectroscopy~\citep{sing19,cubillos20} and exoplanet host star radiation fields~\citep{france16, loyd18a, sparcs21}.  NASA has embraced this opportunity with a dedicated funding line for astrophysics CubeSats (full mission life cycle cost $<$~\$10M) and the Pioneers program (mission cost \$10M~--~\$20M).  In this paper, we present an overview of NASA’s first UV astronomy CubeSat and the first grant-funded small satellite dedicated to the characterization of extrasolar planetary atmospheres, the {\it Colorado Ultraviolet Transit Experiment} (CUTE).   

$CUTE$ conducts transit spectroscopy of short-period, giant planets in the near-UV (2479~--~3306~\AA) bandpass to access strong atomic transitions tracing atmospheric escape and the near-UV spectral slope of giant planet atmospheres that provide constraints on their composition.   This paper presents the science background for and the technical implementation of the mission.  The manuscript is laid out as follows: the scientific motivation for $CUTE$ and its science objectives are presented in Section 2.  Because $CUTE$ is one of the first astronomy missions to be developed in a CubeSat framework, we present a description of the mission development and implementation path in Section 3.  Section 4 presents the instrument design and high-level performance specifications (see also \citealt{fleming18} for a description of $CUTE$’s science payload).   Section 5 describes $CUTE$’s mission operations and we present early-release examples of the mission’s on-orbit science data in Section 6.  We conclude with a brief summary in Section 7.  A detailed description of $CUTE$’s science instrument and on-orbit performance is presented in a companion paper by A. Egan.  Mission operations and on-orbit commissioning (Suresh et al. – in prep), $CUTE$’s on-orbit data pipeline (Sreejith et al. –  in press), and Early Release Science results (Egan et al. – in prep; Sreejith et al. – in prep) will be described in forthcoming papers. 

\begin{figure}
\centering 
\includegraphics[width=1.0\textwidth, angle=90]{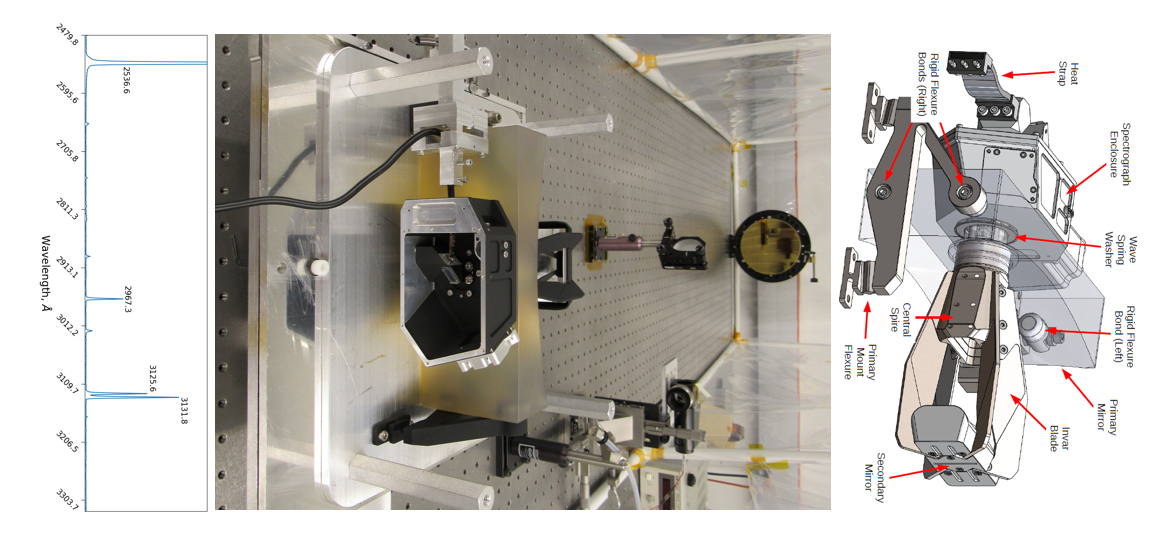}
    \caption{The $CUTE$ instrument development from concept (instrument schematic, top) to telescope characterization ($CUTE$ flight telescope in the test facilities at the University of Colorado, middle), to pre-delivery in-band spectral resolution test data (bottom).     }
    \label{OvsB}
\end{figure}

\section{CUTE SCIENCE OBJECTIVES}
\label{sec:obs}

Planetary escape processes play a key role in determining the chemical and physical state of planets both within and beyond our solar system. Atmospheric escape is thought to create the radius gap observed in the distribution of short-period exoplanets~\citep{fulton18}, likely driven by a combination of photoevaporative~\citep{owen17} and core-powered~\citep{ginzburg18} mass loss.  Escape is also a fundamental process in the evolution of terrestrial worlds.  
For a planet to be habitable, our current view is that it must lose its primordial hydrogen atmosphere and acquire/generate (and retain) a secondary atmosphere~\citep{lammer18}.  Atmospheric escape is known to have shaped the early atmospheres of Venus, Earth, and Mars, which subsequently followed different evolutionary paths.  The rapid hydrodynamic escape that is believed to have affected Venus, Earth and Mars in the past no longer takes place on any planet in the solar system. Therefore, we turn to short-period extrasolar planets as laboratories on which to study vigorous atmospheric loss.  

The first detection of exoplanet atmospheric escape was achieved by \citet{vidal03} who used HI Ly$\alpha$ transit observations in the far-ultraviolet (FUV) to observe the extended atmosphere of the Hot Jupiter HD209458b.  This was followed by the detection of \ion{O}{1}, \ion{C}{2}, \ion{Si}{3} and \ion{Mg}{1} on the same planet \citep{vidal04,vidal13, linsky10}.  These initial observations inspired several independent groups to develop 1D and 3D models to study both the physical characteristics of the upper atmospheres of close-in planets and the escaping gas and plasma surrounding them (e.g.,
~\citealt{koskinen07,murray-clay09, koskinen13a, koskinen13b, bourrier13, bourrier16,dangelo18,carolan21}).


The interpretation of FUV transit measurements has often been controversial (see \citealt{fossati15} for a discussion).  Recently, several atmospheric escape studies have shifted to the near-ultraviolet (NUV), where the stellar flux is much higher than in the FUV and the light curves are measured against a better-understood intensity distribution from the stellar photosphere (e.g.,\citealt{haswell12,llama15}).  The NUV includes the \ion{Fe}{2} complexes near 2400 and 2600 \AA, the \ion{Mg}{2} doublet at 2796/2803 \AA, the \ion{Mg}{1} line at 2852 \AA, some of which have been detected on the Hot Jupiters WASP-12b, HD209458b, and WASP-121b \citep{fossati10,sing19,cubillos20}. 
We note that the \ion{Fe}{2} and \ion{Mg}{2} resonance lines in the near-UV trace the highly extended (and potentially escaping) exoplanet atmosphere, whereas optical band metal line detections made with ground-based telescopes trace the lower, bound atmospheric layers~\citep{hoeijmakers19,casasayas19,cauley19,turner20,hoeijmakers20,casasayas21,deibert21}.  The NUV also contains a pseudo-continuum that can probe scattering by high altitude clouds and gas phase silicon and magnesium~\citep{lothringer22},  as well as the $A$~--~$X$ bands of OH (3100 \AA).  Furthermore, NUV transmission spectra give the unique opportunity to constrain the composition of the aerosols lying in the lower atmospheres~\citep{cubillos20}.
\nocite{haswell12,llama15,vidal13, fossati15}


Depending on the temperature profile in the atmosphere, species like Si, Mg, and Fe are expected to condense to form clouds in the lower atmosphere, however, the calculations indicate that strong mixing, either by turbulence or global circulation, can inhibit cloud formation or allow for these species to be present in the upper atmosphere where they can escape~\citep{koskinen13b,cubillos20,koskinen22}.  The comparison of continuum and atomic line absorption therefore acts as a diagnostic of cloud formation, elemental abundances and mass loss on close-in exoplanets~\citep{lothringer20,cubillos20}.  Model outputs can be used to translate observed planetary transit light curves into global mass-loss rates: the depth and shape of the light curves directly relate to the atmospheric parameters.

Finally, UV transits with $HST$ have provided evidence for time-variability, potentially arising from changing stellar high-energy input, orbital timescale changes in the planet's atmosphere, or variation in the star-planet magnetic environment.  \citet{lecavelier12} observed time-variable neutral hydrogen absorption in FUV transit observations of HD 189733b, possible due to the influence of high-energy stellar flares.  NUV transit observations of the close-in giant planet WASP-12b by \citet{fossati10} found that the transit light curve of WASP-12b presents both an early ingress when compared to its optical transit and excess absorption during the transit (see also \citealt{haswell12,nichols15}).  Possible explanations include atmospheric hydrodynamic mass-loss supporting a shock upstream of the planet’s orbit or generating an accretion stream that produces an early ingress~\citep{lai10,bisikalo13,turner16a}
and a magnetically supported bow-shock 4 – 5 planetary radii upstream of the planet’s orbital motion (analogous to the Earth-Sun system; \citealt{vidotto10,llama11}). 

$CUTE$’s primary science goal is to provide new constraints on the physics and chemistry of hot, Jovian-size exoplanets.  The $CUTE$~ mission addresses this goal with the following observing program: 
\begin{enumerate}
    \item Measure NUV transmission spectra for a small survey of approximately 10 short period planets  
    \item Infer atmospheric escape rates and constrain the composition of the upper atmospheres of hot giant planets
    \item Measure temporal variability in UV transit light curves by observing 6~--~10 transit observations per planet
    \item Measure out-of-transit baseline fluxes to better characterize the stellar inputs to the planet's atmosphere and to capture light curve asymmetries
\end{enumerate}

$CUTE$'s instrument design and mission implementation was developed to enable the four key goals of the observing program.  The spectral coverage and resolution of the $CUTE$ ($\Delta$ $v$ $<$ 300 km s$^{-1}$) spectrograph provides ample separation of the relevant atomic, molecular, and continuum bands in this range (see, e.g., Figure 8 of \citealt{sing19}). $CUTE$'s mission design complements the instrument to meet the science goals of the mission.  (1) We couple observations of the NUV continuum opacity, individual ionic tracers (\ion{Fe}{2}, \ion{Mg}{2}) with atmospheric chemistry, and hydrodynamic escape models to determine mass loss rates for $CUTE$'s targets.  The sample size is driven by a combination of mission lifetime and instrumental sensitivity considerations. (2) $CUTE$ measures the amplitude and slope of the NUV transmission curve to provide constraints on the chemistry and structure of the escaping atmosphere.  The instrumental effective area was specified to enable multiple, wavelength resolved, NUV bands with sufficient photometric precision to distinguish the NUV transit radius from the white-light radius of the planet on all 10 targets and isolate transit spectra of the strongest absorption lines (e.g., \ion{Fe}{2} and \ion{Mg}{2}) on the brightest targets (addressing goals 1 and 2).  The target sample was defined by estimating the detectability of excess NUV absorption; a combination of stellar brightness (V-magnitude), spectral type (A- and F-type stars have spectral energy distributions peaked in the NUV), planetary radius, effective planetary surface temperature, and gravity (hotter, lower-mass planets being more likely to exhibit extended atmospheres).  (3) $CUTE$'s point-stare-repeat concept of operations is designed to make numerous visits to the same planet over the course of 4 to 8 weeks, building signal-to-noise for fainter targets and enabling measurements of light curve variability for brighter targets.  (4) The same point-stare-repeat observing mode provides a wide stellar baseline to measure changes in the \ion{Mg}{2} activity and the increased dispersion of the photospheric and chromospheric continuum flux that indicate variability in the star's escape-driving XUV output.




\section{Mission Implementation Path}
\label{sec:obs}

$CUTE$ is NASA's first grant-funded UV/ Optical/ Infrared small satellite and first dedicated exoplanet spectroscopy mission.  Given the novelty of this mission format for astrophysics science missions, we present a brief overview of the process, schedule, and cost of the mission here.  The initial motivation for $CUTE$ was discussed at a Keck/KISS workshop on exoplanet magnetic fields in August 2013, with the final science and measurement concept in place by the summer of 2015 following numerous informal discussions at science conferences.  Fall 2015 was spent on science measurement definition and the development of the $CUTE$ instrument design. 

$CUTE$ was proposed as a four-year program through NASA's ROSES2015 call (submitted in March 2016), at an initial cost-to-launch of \$3.3M, comparable to an astrophysics sounding rocket proposal but considerably lower cost than a stratospheric balloon program.  $CUTE$ was proposed and selected prior to the initiation of dedicated funding for astrophysics CubeSats, leading to a long delay between proposal submission and the start of funding (approximately 16 months; there was no Phase A or Concept Study period for $CUTE$).  Long lead-time items such as the $CUTE$ spacecraft bus, the rectangular telescope, holographically-ruled diffraction grating, and NUV-optimized CCD detectors were ordered a few months after selection in fall 2017.  The $CUTE$ spacecraft (Blue Canyon Technology, BCT) costs increased relative to the quote provided for the proposal. To accommodate the cost increases in a mission class without reserves, several descopes were implemented, including scaling back the spacecraft's attitude control system to a single star-tracker and eliminating engineering model radios.  In 2018 and 2019, we developed the hardware test facilities that complemented the University of Colorado's existing UV vacuum calibration facilities~\citep{france16c} and conducted component-level characterization (e.g., groove efficiency of the diffraction gratings, trade study of Al vs. Al+MgF$_{2}$ grating coatings). Instrument assembly and characterization, integration into the spacecraft, and pre-delivery environmental testing (e.g., vibration testing, comprehensive performance testing, thermal vacuum testing, etc.) were completed in 2020 and 2021.  The duration from the start of $CUTE$ funding to delivery of the completed observatory was almost exactly four years, although approximately 10 months of schedule were lost to the COVID-19 pandemic.

$CUTE$ proposed to NASA's CubeSat Launch Initiative (CSLI) for launch support in fall 2017 and was selected for flight.   The proposed spacecraft orbit, including the initial $CUTE$ mission requirement documentation submitted to CSLI, requested a dawn-dusk (terminator), sun-synchronous orbit to enable uninterrupted orbital phase coverage of transiting planets and to minimize day/night thermal variations.  Orbital altitude (450 - 600km) was a secondary consideration driven by desired mission lifetime.  CSLI was unable to accommodate the requested sun-synchronous orbit within the time-period covered by the mission funding and $CUTE$ was instead  manifested in November 2019 as a secondary payload on NASA's Landsat 9 mission.  The Landsat 9 launch was scheduled 8 months after $CUTE$'s targeted launch window.  As a result, NASA provided a 6 month, \$0.5M extension to the $CUTE$ program in Fall 2019.  Starting in Spring 2020, COVID delayed the Landsat 9 launch by another 9 months.  COVID also impeded $CUTE$'s development timescale owing to supply-chain delays and the challenges of getting students, scientists, and engineers into $CUTE$'s labs for continued testing and development.  The $CUTE$ mission submitted a follow-on, competed APRA proposal in December 2020 to conduct mission operations and carry out the science program (bringing $CUTE$'s cost to complete the full science mission to approximately \$5M).  Of the 18 CubeSat missions originally manifested with Landsat 9, only four (including $CUTE$) would ultimately deliver and be flown on the mission.  

The integrated and tested observatory was delivered to NASA at Vandenberg Space Force Base (VSFB) in July 2021, and the $CUTE$ team supported installation into the CubeSat dispenser on the ESPA ring.  The mission was launched on September 27 2021 into a sun-synchronous orbit ($\approx$98\arcdeg, 10am Local Time of Ascending Node; LTAN) with a 560km apogee.  $CUTE$ deployed from the dispenser approximately two hours after launch; solar arrays deployed and the communication beacon started approximately 30 minutes later.  $CUTE$'s beacons were identified by the amateur RF community on the first orbit and communications were established with the ground station at the University of Colorado on September 28 2021. We refer the reader to Section 4 and Suresh et al. (in preparation) for a description of the $CUTE$ ground segment and commissioning program.  We refer the reader to the $SmallSat$ Conference proceeding by \citealt{egan22_smallsat} for a discussion of lessons learned during $CUTE$ development and early on-orbit operations.

$CUTE$ is part of NASA's suborbital program, where student training and early-career mentorship are key ingredients to the definition of mission success.  $CUTE$'s approach was built off of the framework of the NASA Sounding Rocket Program, which has a long history in the professional development of NASA's space scientist workforce.  The core science and instrument team (defined as those working with $CUTE$ for more than 2 years of the implementation phase) included two Ph.D. candidates (in astrophysics and aerospace engineering), four undergraduates, two postdoctoral researchers, two early career engineers ($CUTE$ was the first job post-bachelors degree for the mission's lead mechanical and electrical engineers), and the early-career project scientist (Dr. Brian Fleming, who became the PI of a NASA sounding rocket program and the SPRITE CubeSat~\citep{fleming22} mission during the course of the $CUTE$ development phase). Over the course of $CUTE$'s component-level and instrument test phase, the project employed another six undergraduate students in various laboratory and science program development tasks (e.g., target field checking for crowded fields).  In addition to this, the operations team for $CUTE$ (see Section 5) included an additional two undergraduate students, two graduate students, and one flight software undergraduate student.  Taken as a whole, $CUTE$ supported the mentoring and training for over 20 early-career scientists and engineers through the completion of the on-orbit commissioning phase.    

\begin{figure}[htbp]
   \centering
   \includegraphics[scale=.44,clip,trim=0mm 0mm 5mm 5mm,angle=0]{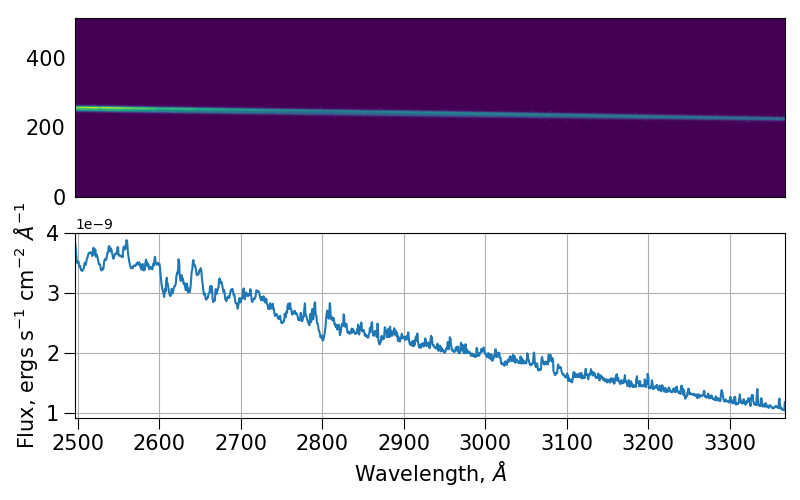}

   \figcaption{$CUTE$ calibration observations of the 
   O4 supergiant $\zeta$ Puppis. The top plot shows the full-frame (2048~$\times$~515 pixel) calibration image and the bottom plot shows the extracted one-dimensional spectrum (flux calibrated against archival $HST$ and $IUE$ spectra).   $\zeta$ Puppis was selected as a calibration target because of the wealth of archival NUV spectra and the high photospheric temperature ensures that iron and magnesium lines in the spectrum are narrow, interstellar features.  }

\end{figure}

\section{Implementation: The $CUTE$~ Science Payload}

The $CUTE$ payload is a magnifying NUV spectrograph fed by a rectangular Cassegrain telescope. The spectrogram is recorded on a back-illuminated, UV-enhanced e2v CCD42-10 that is maintained at a nominal operating temperature ($-$15~--~$-$5\arcdeg\ C) by passive cooling through a radiator panel.  $CUTE$ employs the BCT XB1 bus to provide critical subsystems including power, command and data handling, communications, and attitude control (ADCS).   Figure 1 shows an instrument schematic, optical testing of the telescope, and an in-band calibration spectrum from the flight instrument prior to instrument integration into the BCT chassis.  

The $CUTE$ instrument is housed in 4U of the 6U spacecraft. The $CUTE$ aperture is a 206 $\times$ 84 mm, f/0.75 (in the cross-dispersion axis) primary mirror that is part of the f/2.6 Cassegrain telescope. The rectangular shape of the primary is  matched to the long axis of the 6U CubeSat chassis and allows for 3 times more throughput than a 1U circular aperture~\citep{fleming18}. The hyperbolic secondary mirror is cantilevered off of the primary mirror, which serves as the bench for the optical instrument, by means of an Invar tower (see Figure 1). A 15 $\times$ 6 mm fold mirror redirects the beam 90\arcdeg\ through a 141 $\mu$m $\times$ 3.5 mm (60\arcsec\ $\times$ 1400\arcsec\ projected) slit at the Cassegrain focus.  The slit, manufactured by OSH Stencils, was polished on the incident side and angled 45\arcdeg\ about the slit axis to redirect the field to an aspect camera for use in telescope performance testing and alignment with the BCT spacecraft during integration.  The rectangular telescope design optimizes collecting area within the mass-volume constraints of the cubesat form factor, while the large sky field-of-view, increased cost, and mechanical stress at the primary mirror-secondary tower interface add design complications.

Once through the slit, the starlight is diffracted, redirected, and magnified by a spherical, R = 86.1mm radius, 1714 gr mm$^{-1}$ aberration correcting, ion-etched holographic grating fabricated by Horiba Jobin-Yvon (Horiba J-Y). The holographic grating design was adopted to minimize scattered light in the system. A second fold mirror with an R$_{x}$ = 300 mm radius of curvature about the cross-dispersion dimension provides additional aberration corrections before the beam reaches the CCD. The final beam focal ratio is f/5.5 in the cross-dispersion axis, with a detector plate scale of 186\arcsec\ mm$^{-1}$.  The detector and custom avionics were tested and flight ruggedized by the $CUTE$ team in their on-campus laboratories at the University of Colorado~\citep{nell21}.  

The telescope was delivered fully assembled to CU by Nu-Tek Precision Optical. All mirrors are coated with MgF$_{2}$ + Al to prevent the formation of an oxide layer (AlO$_{3}$). We elected to receive flight and flight-backup gratings coated in bare Al and MgF$_{2}$ + Al, respectively (coated by Horiba J-Y), to control for a potential efficiency anomaly similar to that seen on the COS NUV gratings~\citep{wilkinson02}.  Detailed pre-flight efficiency and environmental testing showed better performance with the bare Al grating, without a measurable loss in efficiency over time. As a result, the instrument team elected to fly the bare Al-coated mirror on the flight instrument.  The design-prediction flight instrument performance curves are presented in \citet{fleming18} and the on-orbit instrument performance of the $CUTE$ payload is presented in Egan et al. (2022~--~this volume); we provide a brief summary of the key performance metrics in the following subsection.

\begin{deluxetable}{lc}
\tablewidth{0pt}
\tablecaption{$CUTE$ Instrument Specifications\label{tab:modes}}
\tablehead{\colhead{Instrument Metric} & \colhead{On-orbit Value} }
\tabletypesize{\normalsize}
\startdata
Bandpass & 2479~--~3306 \AA\  \\
Spectral Resolution$^{a}$ & 3.9~\AA\ \\
Cross-Dispersion Resolution$^{b}$ & $\approx$~30\arcsec\ \\
Peak A$_{eff}$ &  27.5 cm$^{2}$ at 2500~\AA\ \\
Background Flux Limit & 5~$\times$~10$^{-14}$  erg cm$^{-2}$ s$^{-1}$ \AA$^{-1}$  \\
~~~in 300s$^{b}$ & 
\enddata
\tablenotetext{a}{Average resolution over the bandpass, including spacecraft jitter.}
\tablenotetext{b}{Evaluated at 3000~\AA.}
\vspace{-0.25in}
\end{deluxetable}

\begin{figure*}[htpb]
   \centering
   \includegraphics[scale=.50,clip,trim=0mm 0mm 0mm 0mm,angle=0]{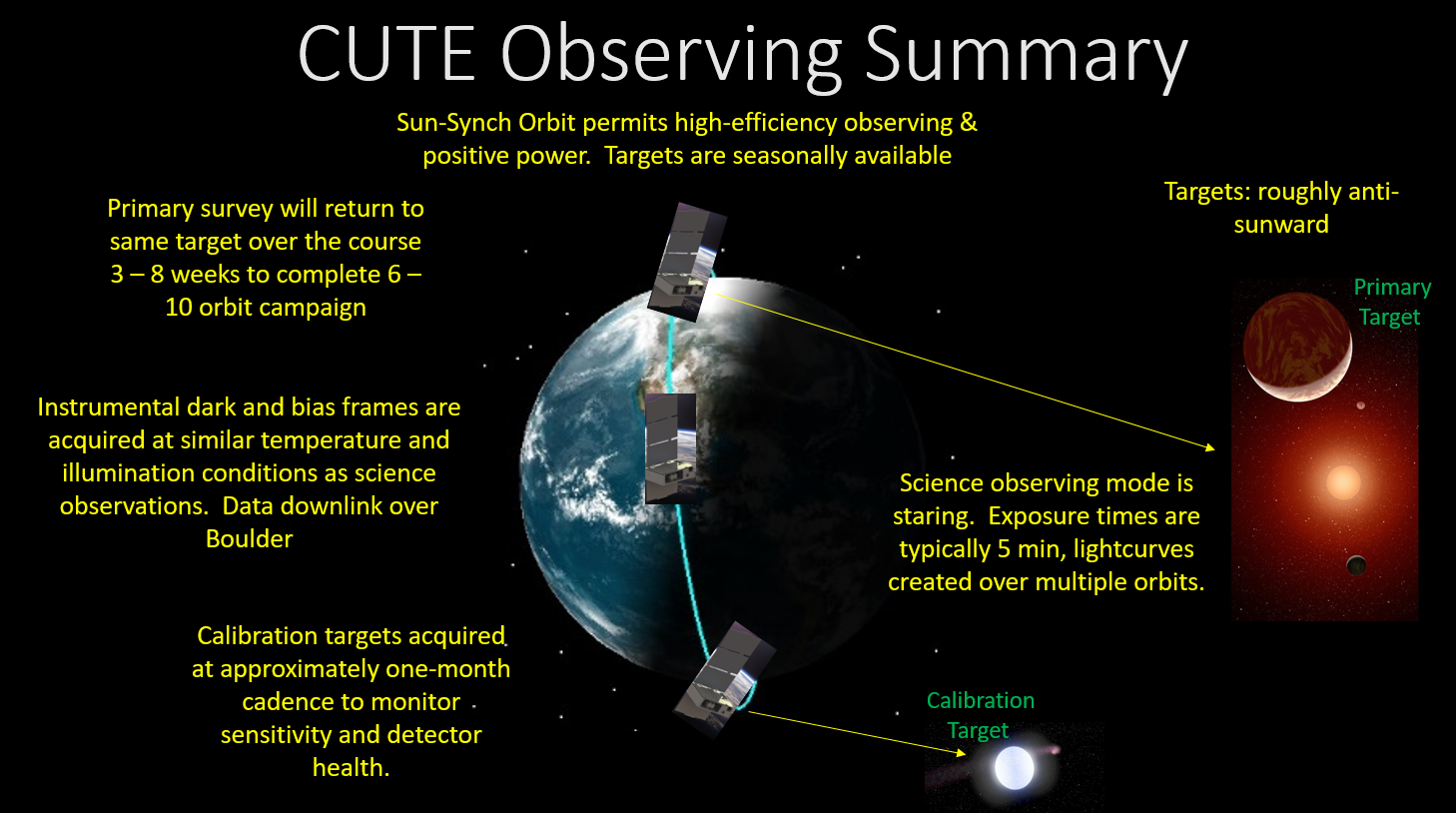}
   \figcaption{Schematic description of $CUTE$ science and calibration observations.
   }
\end{figure*}

\subsection{Instrument Specifications}

The final bandpass recorded by the CCD detector is 2479~--~3306~\AA\ (see Table 1), which is a slight change from the pre-flight projection owing to shifts in the optical system during ascent.  The exact bandpass also varies by several \AA\ depending on the alignment of the stellar point spread function (PSF) in the spectrograph slit. The spectral resolving power of the instrument is $\approx$~750 ($\Delta \lambda$~$\approx$~3.3~--~4.5~\AA\ across the bandpass, including the effects of spacecraft pointing jitter).  Figure 2 shows a representative calibration spectrum from the $CUTE$ on-orbit commissioning program.

The system effective  area is a function of the reflection efficiency of the optics ($R$), efficiency of the grating ($\epsilon_g$), and quantum efficiency of the detector (DQE), multiplied by the geometric collecting area of the telescope (A$_{geo}$):
\begin{equation}
    A_{eff} (\lambda)=A_{geo}  R^{5}(\lambda) \epsilon_g (\lambda) DQE (\lambda).
\end{equation}
The on-orbit effective area was measured by comparing $CUTE$'s observations (in units of electrons s$^{-1}$ \AA$^{-1}$) with flux-calibrated observations from the $IUE$ and $HST$ archives. We measured $A_{eff}$~=~27.5~--~19.0 cm$^{2}$ across the $CUTE$ spectral range with a peak at approximately 2500~\AA.  Component-level efficiencies were measured prior to instrument assembly in the UV calibration facilities at the University of Colorado~\citep{france16c,egan20}. The component-based, pre-flight $A_{eff}$ estimate was about  12\% higher than the median effective area subsequently measured on-orbit  (Egan et al. -- this volume).  We attribute the loss of sensitivity to two possible causes:  particulate contamination during the failure of $CUTE$'s thermal-electric cooling system (which occurred during thermal vacuum testing) and contamination during the $\sim$2 months that $CUTE$ sat in the CubeSat dispenser at VSFB prior to launch.  A dry nitrogen purge was requested in order to minimize optical degradation following dispenser integration, but was not made available.  The difference in effective area does not have a significant impact on target selection and detectability, however, the larger and variable thermal environment resulting from the loss of the active cooling system removes most of the stars fainter than the nominal target list. 

Combining the $CUTE$ effective area with the on-orbit instrumental background level and the nominal 300 second exposure time for $CUTE$'s exoplanet surveys, we calculate the typical dispersion in the residual flux following background subtraction.  This sets the minimum flux level that can be detected above the noise in a 300 second spectrum, which we refer to as the background flux limit.  We measure a background flux limit of $\approx$~5~$\times$~10$^{-14}$  erg cm$^{-2}$ s$^{-1}$ \AA$^{-1}$ at 3000~\AA.

\section{$CUTE$ Mission Operations}

The $CUTE$ spacecraft includes a UHF (437.25 MHz) antenna with both transmission and receiving capabilities.  The UHF link is used for uploading commands to the spacecraft and monitoring real-time telemetry during ground passes. $CUTE$ also has an S-band (2402 MHz) downlink-only mode for primary science data transmission.  
The mission operations and ground station for $CUTE$ are located at the Laboratory for Atmospheric and Space Physics in Boulder, Colorado. $CUTE$ typically has 1~--~2 high-elevation ($>$~50\arcdeg) passes and 1~--~2 low-elevation passes per day over the Boulder ground station, resulting in approximately 10 minutes per day of optimal downlink time.  Figure 3 presents an illustration of the $CUTE$ science operations observing mode, including science data acquisition, approximately monthly calibration activities, and data downlinks over the Boulder ground station.  

The CubeSat Operations Center at LASP utilizes the LASP ground station initially built for CSSWE and MinXSS, and the recently completed CSIM CubeSat mission for NASA's heliophysics division~\citep{mason16}, using a combination of HYDRA and OASIS-CC~\citep{flynn21} for command and control. The mission operations are conducted by a team of professionals with experience from larger NASA flight missions (e.g., $Kepler$, $IXPE$, and several heliophysics missions) and a dedicated student operations group.  The undergraduate and graduate student operators perform mission planning, operations, and health and status monitoring for the spacecraft; the science team has developed a graphical user interface tool to determine optimal target visibility and viewing conditions. The output of the science planning tool is processed into $CUTE$'s weekly operations plan to define the observing, charging, and communication activities for the week. 

 To maximize operational simplicity, $CUTE$ conducts single-target campaigns: we schedule multiple transit observations in a single command block (typically lasting 3 - 7 days) that is uploaded to the spacecraft.  Each command block includes calibration exposures, science exposures, and data downlink periods.  We repeat this exercise until 6~--~10 transits of a given planet have been executed.  Interleaved with the transit observing blocks are dedicated downlink periods, typically between 3 and 5 days, to re-transmit science and calibration data that experienced low data-completion fractions in the initial downlink or were lost to spacecraft resets.  The need to execute 1~--~3 additional data downlinks per transit campaign, driven by the frequent loss of fine-pointing control and resets on the spacecraft bus (1~--~2 events per week), is the limiting factor to $CUTE$'s operational efficiency.  Pre-flight observation planning  predicted 3 transits per week could be successfully executed and downlinked, which would complete 10 transits per target of the 10 target sample in approximately 8 months.  The realized mission efficiency is projected to complete an average of 6 transits per target over a science mission lifetime of $\sim$~15 months, or approximately a factor of 3 reduction in efficiency compared to pre-flight estimates.   A summary of the $CUTE$ on-orbit operations and payload commissioning will be presented in a forthcoming paper (A. Suresh et al. - in prep.).

\begin{figure*}[htbp]
   \centering
   \includegraphics[scale=0.52,clip,trim=0mm 0mm 0mm 0mm,angle=0]{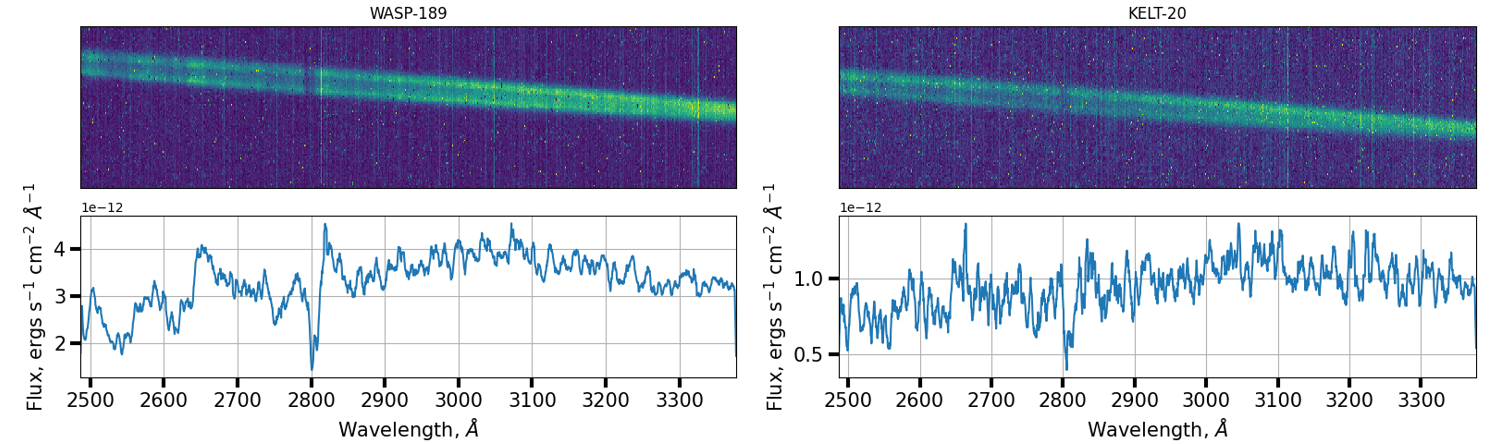}

   \figcaption{$CUTE$ spectral observations of WASP-189 and KELT-20, A-type host stars of $CUTE$'s first Early Release Science targets. The top plots shows a "TRIM2D" 2048~$\times$~100 pixel two-dimensional data product (exposure time = 5 minutes).  The bottom plots show the one-dimensional spectral collapse including wavelength and flux calibration. A 10 pixel boxcar smooth has been applied to the one-dimensional spectra for display purposes. }
\end{figure*}

\section{$CUTE$ Science Data Example}

  In this section we present representative samples of $CUTE$'s individual science data products and a preliminary reduced transit light curve from the Early Release Science program.  Detailed analyses of the wavelength dependent transit depths, interpretation and quantification of atmospheric composition and escape rates, and inter-comparison between different transit visits for the Early Release Data program will be presented in upcoming works (Egan et al. and Sreejith et al. 2022~--~in prep).  The goal here is to present flight data of exoplanets and their host stars to illustrate the features (and limitations) of $CUTE$ observations as revealed by the first two targets of the Early Release Science program.

Figure 4 ($top$) presents a standard spectral data product that $CUTE$ transmits over the S-band downlink.  These are ``TRIM2D'' data products, 2048~$\times$~100 pixel two-dimensional spectra with a 5 minute exposure time.  The images are trimmed to reduce downlink volume.  The 5 minute exposure time is typical of all $CUTE$ spectra and is a balance of signal-to-noise for our target brightness, number of exposures possible per orbital night, and simplicity of operational planning.  Each transit visit is buffered by a number of bias and dark exposures taken at similar celestial pointing, orbital position (latitude and longitude, and therefore similar temperature and illumination conditions), and elevation angle of the telescope with respect to the Earth limb.  These calibration files are used to remove thermal and readout noise effects, as described in Egan et al. (2022).  

Data processing beyond the downlink of the TRIM2D data products occurs on the ground.  The two-dimensional data are collapsed along a diagonal extraction region; the spectra are wavelength and flux calibrated using observations from the on-orbit commissioning phase. Figure 4 ($bottom$) shows calibrated one-dimensional spectra of WASP-189 and KELT-20, taken outside of transit.  The spectra are typical of NUV observations from A-type stars in the $IUE$ archive, with the most prominent feature being \ion{Mg}{2} absorption in the photosphere of these intermediate temperature stars.  The reader will also notice the defocus seen in the cross-dispersion direction, manifest as the double-lobe structure that increases to shorter wavelengths across the band.  This defocus was introduced during an additional payload vibration test that was not part of the original test specifications but later required by NASA just prior to the delivery of the spacecraft.  This defocus was then exacerbated during the powered ascent; there is no detectable ``breathing'' of the focus with orbital location although the background levels are strongly driven by the spacecraft thermal and illumination conditions.

$CUTE$'s two-dimensional spectra are calibrated with master bias frames before cosmic ray correction using LAcosmic algorithm \citep{dokkum}. We extract the one-dimensional spectra from this corrected image as described in Sreejith et al. (2022~--~in prep). The one-dimensional spectra are obtained by subtracting the background, which are then wavelength calibrated. The spectra are integrated over the wavelength region of $\approx$2540\AA\ to $\approx$3300\AA\ to create a light curve point. 
These light curves can be created down to a wavelength resolution limit of $\sim$~4~\AA\ per bin, but for initial demonstration purposes we display a broad NUV bandpass.

\begin{figure*}[htbp]
   \centering
   \includegraphics[scale=.42,clip,trim=0mm 0mm 5mm 5mm,angle=00]{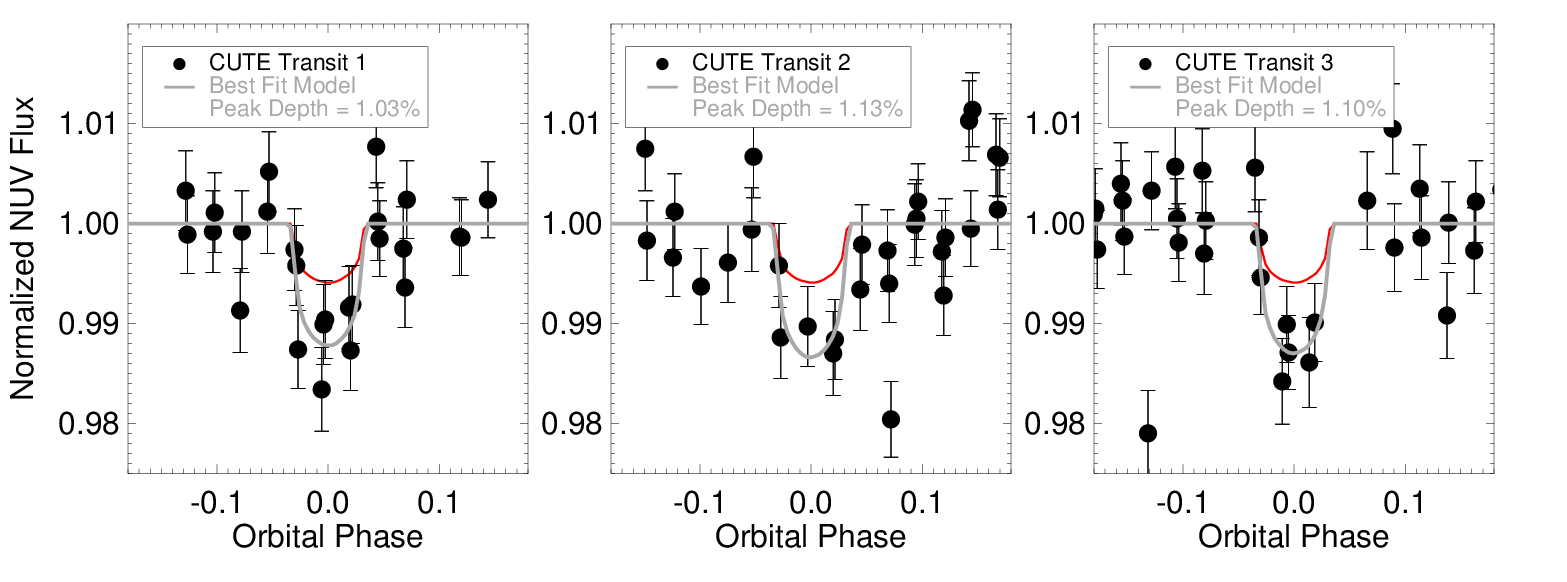}

   \figcaption{Initial $CUTE$ light curves of WASP-189b, showing three independent NUV (approximately 2540~--~3300~\AA) light curves (black points) and the best-fit transit models in gray.  The plots compare the NUV band light curves with the optical light curve (in red) from $CHEOPS$~\citep{lendl20}.  The NUV transits are significantly deeper than their broadband optical counterparts, indicating an effective planetary radius increase of R$_{P,NUV}$~$\approx$~1.5 R$_{P,opt}$.   }
\end{figure*}

Figure 5 presents $CUTE$'s approximately 2540~--~3300~\AA\ light curves for 3 different visits of the ultra-hot Jupiter WASP-189b.  The best fit transit model, taking into account wavelength-dependent stellar limb darkening (shown in gray), will be presented in detail in Sreejith et al. (2022~--~in prep).  The optical transit light curve from $CHEOPS$~\citep{lendl20} is shown for comparison and suggests excess transit absorption at UV wavelength compared with the broadband geometric size of the planet.   We demonstrate self-consistent transit depth recoveries of $\approx$~1.0~--~1.1~\% over three separate transit observations of WASP-189b separated by several weeks. 

Excess planetary absorption at NUV wavelengths is consistent with previous observations of ultra-hot Jupiters observed with $HST$~\citep{sing19,cubillos20,lothringer22}.   A transit depth of 1\% in WASP-189b would indicate that the NUV transit observations are probing the extended upper atmosphere of the planet that is subject to stellar high energy radiation and escape. This is because the 1 microbar level, used here as a rough proxy for the base of the thermosphere, has a radius of about 1.1 R$_{p}$ and a transit depth of 0.6\%, based on an effective temperature of 2410 K and an atmosphere with solar abundances. In contrast, a transit depth of 1\% corresponds to a larger radius of 1.4 R$_{p}$. If we assume a temperature of 8000 K in the thermosphere, the pressure at 1.4 R$_{p}$ would be between 0.1 and 1 nbar. Given that this pressure is too low for significant clouds and hazes, a pseudo-continuum by these absorbers is unlikely and the broadband NUV transit depth likely arises from a forest of metal ion lines (e.g., \citealt{fossati10,sing19}). The individual absorption lines responsible would have to extend to much higher radii than 1.4 R$_{p}$ in transit to be detectable in the broad NUV band.   


 We note that the preliminary lightcurves show significant scatter beyond the photon noise limit at this stage of the reduction.  Work is ongoing to model the temperature- and orbital position-dependent background to reduce the observed dispersion in the lightcurves.

\section{Conclusions}
\label{sec:conc}

The $CUTE$ CubeSat mission was launched in September 2021 and is currently carrying out its primary science mission to collect NUV spectroscopy of transiting planets.  $CUTE$ has successfully completed spacecraft and instrument commissioning and has completed initial science observations on a number of exoplanetary systems.  Optimal science targets are short-period, Jovian-sized planets orbiting bright ($V$~$<$~8) F and A stars. The science instrument has demonstrated sensitivity to NUV white light transit depth of $<$~1\% and wavelength-dependent exoplanet atmosphere opacity increase at $\lambda$~$\lesssim$~3300~\AA.   

We have presented the motivation for, mission design of, and on-orbit characteristics of the $CUTE$ mission.  The companion paper by Egan et al. present the details of the on-orbit instrument performance of $CUTE$ and future papers will present mission science results as well as information about the $CUTE$ ground-segment data pipeline, commissioning, and operations of the mission. $CUTE$ has supported mentoring and training for over 20 early-career scientists and engineers.  Mission operations are planned to continue until June 2023  (currently limited by project funding) and the initial release of $CUTE$ data will be delivered to the NexSci archive in 2023.

 Acknowledgments: $CUTE$ was developed and operated with the support to two NASA/APRA awards to the Laboratory for Atmospheric and Space Physics at the University of Colorado Boulder, NNX17AI84G and 80NSSC21K1667.  A. G. S. was supported by a Schr\"{o}dinger Fellowship through the Austrian Science Fund (FWF) [J 4596-N]. A. G. S. and L.F. acknowledge financial support from the Austrian Forschungsf\"orderungsgesellschaft FFG project 859718 and 865968. 
AAV acknowledges funding from the European Research Council (ERC) under the European Union's Horizon 2020 research and innovation programme (grant agreement No 817540, ASTROFLOW). 
K.F.  acknowledges the numerous and invaluable discussions with colleagues excited about ultraviolet transit science and the potential to do science with small satellites. The $CUTE$ team wishes to specifically recognize the amateur radio operator community and, and SatNOGS network specifically, for hosting numerous telemetry tracking tools that have improved the mission's ability to recover from faults and understand long-term spacecraft trends much more efficiently than would have been otherwise possible.

\clearpage

\bibliography{cute_bibtex}{}
\bibliographystyle{aasjournal}

\end{document}